\documentclass[prd, reprint, preprintnumbers, twocolumn, eqsecnum,floatfix,letterpaper,superscriptaddress,nofootinbib, wrapfig, caption]{revtex4-1}
\usepackage{graphicx,url,amssymb,amsmath,rotating,color,units,wasysym,epsfig,multirow,epstopdf}
\usepackage[colorlinks,urlcolor=blue,citecolor=blue,linkcolor=blue]{hyperref}
\usepackage{comment}
\usepackage{float}
\usepackage{graphicx}
\usepackage{subcaption}
\usepackage{wrapfig}
\usepackage{ulem}

\usepackage{booktabs,tabularx,array,makecell}
\newcolumntype{Y}{>{\raggedright\arraybackslash}X}
\newcolumntype{L}[1]{>{\raggedright\arraybackslash}p{#1}}

\graphicspath{{./}{figures_v3/}}

\begin{document}


\title{The best of all worlds: gravitational-wave transient detection with multiple pipelines}

\author{Nikolas Moustakidis}
\affiliation{Department of Informatics, Aristotle University of Thessaloniki, Greece}
\affiliation{MIT-LIGO Laboratory, 185 Albany St, MIT, Cambridge, MA 02139, USA}
\affiliation{Institute for Accelerated AI Algorithms for Data-Driven Discovery (A3D3), MIT, Cambridge, MA 02139, USA}
\author{Theofilos Moustakidis}
\affiliation{Department of Bioinformatics, University of Thessaly, Greece}
\affiliation{MIT-LIGO Laboratory, 185 Albany St, MIT, Cambridge, MA 02139, USA}
\affiliation{Institute for Accelerated AI Algorithms for Data-Driven Discovery (A3D3), MIT, Cambridge, MA 02139, USA}
\author{Deep Chatterjee}
\affiliation{MIT-LIGO Laboratory, 185 Albany St, MIT, Cambridge, MA 02139, USA}
\affiliation{Institute for Accelerated AI Algorithms for Data-Driven Discovery (A3D3), MIT, Cambridge, MA 02139, USA}
\author{Anastasios Tefas}
\affiliation{Department of Informatics, Aristotle University of Thessaloniki, Greece}
\author{Erik Katsavounidis}
\affiliation{MIT-LIGO Laboratory, 185 Albany St, MIT, Cambridge, MA 02139, USA}
\affiliation{Institute for Accelerated AI Algorithms for Data-Driven Discovery (A3D3), MIT, Cambridge, MA 02139, USA}

\date{\today}


\title{The best of all worlds: gravitational-wave transient detection with multiple pipelines}

\begin{abstract}
The search for gravitational-wave transients (modeled and unmodeled)
as performed by the LIGO-Virgo-KAGRA collaborations and the broader community typically involves multiple detection algorithms, often redundant and sometimes complementary.
We here address the problem of combining such detection algorithms in order to maximize the noise vs signal discrimination power of such an astrophysical search that utilizes multiple analysis pipelines.
Using a machine learning approach, we construct a neural network-based classifier that utilizes information (or lack thereof) from all participating pipelines in the search for compact binary coalescences in LIGO-Virgo-KAGRA's real-time (and offline) processing of gravitational-wave data. 
We benchmark the method with simulated (astrophysical) signals and real noise from the instruments and obtain receiver operating characteristic curves to quantify its performance.
We show how this leads to an effectively new, combined pipeline
that outperforms any single pipeline alone or other logical combinations of them.
This is particularly critical for real-time operations of the LIGO-Virgo-KAGRA detectors as such combination of detection methods can lead to increased robustness in identifying astrophysical sources and to reduction of the number of false alerts that may otherwise be sent out as astronomical telegrams.
Our method also aims in simplifying the presentation of astrophysical results out of the LIGO-Virgo-KAGRA detectors by combining all participating pipelines in the astrophysical searches as well as optimizing resources by the broader multi-messenger astrophysics community in following up astronomical alerts for gravitational-wave transient events.
\end{abstract}
 \maketitle

\section{Introduction} \label{sec:intro}

The ground-based km-scale laser interferometers LIGO~\cite{Aasi_2015} and Virgo~\cite{Acernese_2015} have established the field of gravitational-wave (GW) astronomy with direct observations of astrophysical sources of gravitational radiation over close to a decade now~\cite{AbEA2016a,Abbott_2023,theligoscientificcollaboration2025gwtc40updatinggravitationalwavetransient}.
Together with the KAGRA interferometer~\cite{kagra} they form the LIGO-Virgo-KAGRA international network of detectors and corresponding collaborations (referred to as LVK) responsible --among other things-- for the analysis of the data such network collects and the publishing of catalogs of GW transient sources that they detect.
LVK completed in late 2025 (November 18) observing the GW sky in its fourth observing run (O4) with new detections of transient events now being a common occurrence, about once every few days~\cite{Abac_2025}.

All GW sources reported so far correspond to mergers of compact binary systems made up of any combination of (stellar mass) black holes and neutron stars-- these sources are collectively referred to as compact binary coalescences, or CBCs.
Signals from binary systems that include only black holes (referred to as binary black holes, or, BBHs) typically last on the order of 1s within the instruments' sensitive frequency band~\cite{Abac_2025}.
Systems involving neutron star(s) will result to longer in-band signals and up to 100s. This includes neutron star-black hole (NSBH) and binary neutron star (BNS) systems~\cite{Abac_2025}.
Given their duration, these are generally considered to be transient sources and the signal-processing algorithms for their detection correspondingly invoke transient-finding techniques.
CBC signals have the advantage of being well modeled; this enables the use of matched-filtering~\cite{matchedfilter} for their optimal detection~\cite{guide_to_lvk_data}.
With the onset of machine learning (ML) techniques, CBC searches are equivalently pursued using fully-supervised ML approaches (for example~\cite{PhysRevD.111.042010,1v7r-bkzs,PhysRevD.103.102003,ares_gw}).
Contrary to the CBC signals, a number of astrophysical processes and systems may give rise to short-lived GW events that are either poorly modeled or not modeled at all.
The search for such unmodeled transients involves general excess power techniques combined with multi-sensor coherent analyses 
(\cite{wjdz-jdby,83j3-pgk1} and references therein), or in the case of ML algorithms, semi-supervised or unsupervised approaches (for example~\cite{Raikman_2024,zykf-8klg,mly}).

Real-time transient searches within the LVK are among the most crucial analysis activities, as they enable the prompt identification of astrophysical events that can be rapidly followed up across the electromagnetic spectrum and in neutrinos. This capability made possible the multi-messenger observations of the first binary neutron star event, GW170817, and the wealth of astrophysical insights it provided~\cite{PhysRevLett.119.161101,Abbott_2017}. LVK data analysis employs multiple search pipelines targeting both modeled (i.e., CBC) and unmodeled transients~\cite{Abac_2025,wjdz-jdby,83j3-pgk1}, supporting the issuance of real-time public alerts as well as the publication of observational results in event catalogs following months-long observing runs~\cite{O2_LL_paper,PhysRevX.13.041039,Abac_2025}.

The logical ``OR" of such pipelines is generally being used for the adaption of a transient source either in low latency or in catalog publications~\cite{Abbott_2023,theligoscientificcollaboration2025gwtc40updatinggravitationalwavetransient,USERGUIDE}.
Search pipeline consistency is a challenge in the presence of multiple algorithms looking over the same dataset, and in the recent catalog, several candidates are reported by a single search pipeline only (for example, see Sec. 2.1.3 in ~\cite{theligoscientificcollaboration2025gwtc40updatinggravitationalwavetransient}). This is a limitation, and a rigorous treatment to quantify the effect of a trials factor is lacking.
Research into how to combine results from multiple search pipelines commenced during the analysis of data from the initial LIGO and Virgo detectors.
The effort focused mostly in how to construct upper limits for transient sources given the multiple pipelines that were used in the analysis of the initial detector data, but also into how to create a unified statistic for ranking events~\cite{Sutton_2009,PhysRevD.85.122009}.
More recently, an approach was developed~\cite{PhysRevD.108.083043} in order to combine the $p_{astro}$ values (i.e., the probability for a GW CBC event to be of astrophysical origin) as provided by multiple search pipelines.
In our work here we present an ML-based approach to combine the multiple search pipelines used by LVK in the search for CBCs.
We are motivated primarily by the low latency operations of the GW instruments and the need to provide real-time public astronomical alerts corresponding to the detection of GW transients.
This is in order to enable their multi-messenger follow-up.
While we carried out this work for CBC search pipelines, the method can be straightforwardly extended for unmodeled (generic) burst searches, too.
Similar ML-approaches for combining pipelines were taken recently in work reported elsewhere~\cite{Tsukamoto_2026,yfb3-fgf2} for the purpose of unifying the detection statistics, although with different neural network architectures and different feature vector choices.
This paper is organized as follows.
In section~\ref{sec:cbc-algos} we discuss the low latency CBC searches that we consider in this analysis. This is followed by section~\ref{sec:data} where we present the corresponding data that they provide for developing our method.
Sections~\ref{sec:boaw} and~\ref{sec:training} present the ML architecture that we employ and how it is trained.
Results are shown in section~\ref{sec:results} and we conclude with section~\ref{sec:conclusion}.

\section{Low Latency CBC searches} \label{sec:cbc-algos}

For this analysis we consider four pipelines sensitive to CBC sources: GstLAL~\cite{PhysRevD.109.042008,joshi2025newmethodsofflinegstlal}, MBTA~\cite{Allene_2025}, PyCBC~\cite{DalCanton_2021,Nitz_2017} and cWB-BBH~\cite{PhysRevD.105.083018,PhysRevD.93.042004}. These pipelines were selected because they participated with close to 100\% duty cycle in a recent LVK real-time Mock-Data-Challenge (MDC)~\cite{MDCPAPER}.

The first three are based on matched filtering and retain sensitivity to a broad range of component masses for CBC sources covering systems from one solar mass to hundreds of solar masses. The cWB-BBH search pipeline is based on a general-purpose short-duration transient-finding algorithm with mild signal constraints to reflect expectations for a BBH source. Because of that, this search pipeline does not present appreciable sensitivity to BNS and NSBH systems due to their longer signal duration. The reader is referred to publications of the respective pipeline analysis teams for more algorithmic details. For the purpose of our method in combining such pipelines, the inner workings of each one of them are not relevant.

GW interferometers record data in short segments that are calibrated in low latency for downstream physics analyses~\cite{theligoscientificcollaboration2026gwtc40methodsidentifyingcharacterizing}.
All four aforementioned pipelines ingest and process calibrated data in real time, generating candidate events corresponding to instances that deviate from the background distribution and are assigned a false-alarm-rate (FAR). This measures the expectation a given candidate event might have resulted from noise fluctuations (i.e., in the absence of astrophysical signals). Candidate events and features extracted from the searches are packaged and uploaded in a central event database called GraceDB~\footnote{https://gracedb.ligo.org/} on a per pipeline basis. Individual events are then grouped based on a 1s-wide time window to form an entity referred to as a ``superevent", which is meant to capture a single physical transient event in the GW detectors~\cite{USERGUIDE}.

Within GraceDB, all information from individual pipeline events is stored under each superevent. The extracted features for each event that we will utilize in our analysis include the FAR, combined (across the interferometers) signal-to-noise ratio (SNR), and chirp mass, along with the requirement that at least one LIGO detector contributes to the trigger.
We do not use any information from the Virgo detector.
Whenever a pipeline submits more than one event into GraceDB for the same physical event in the detector, we use the one with the highest SNR. Within LVK's processing of multi-detector events and the formation of a superevent, the latter inherits the data products of the individual event that has the highest SNR. The generation of public alerts out of LVK's low latency CBC searches is then based on the FAR of such superevent~\cite{USERGUIDE}.

The essence of our approach in combining multiple pipelines is to utilize the entire vector of features extracted from {\it{all}} searches that contributed individual events in a given superevent (i.e., a physical transient in the detectors) rather than relying solely on the highest SNR event as currently is the case with LVK's public alerts. Our combined pipeline constructs a new scalar quantity that can function as the detection statistic and, as we will demonstrate, offers improved discrimination power between signal and noise. In addition, the pipeline can provide supplementary statistics related to the classification of a superevent, including uncertainties, individual model scores, alternative voting methods (hard or soft), and diagnostic flags. We achieve this by training ML models on a population of simulated CBC sources and on noise instances, forcing our model to learn not only how to discriminate these two classes but also how different search pipelines respond (slightly, or sometimes notably) differently to noise and signal.

\section{Data} \label{sec:data}

Our training dataset involves signals and noise candidates. The signal candidates are based on an extensive MDC that involves thousands of simulated events of CBC and which was used by the LVK to benchmark the real-time alert infrastructure in O4~\cite{MDCPAPER}. This involved the replay of five weeks of instrument data from the third LVK observing run (O3) that were replayed in a way that mimics the real-time collection of data from the instruments. Astrophysical signals corresponding to CBCs were added onto such replayed data via software so that the entire low latency infrastructure with first and foremost the very detection algorithms could process them for the purpose of establishing overall performance.
As part of this data replay and MDC, search pipelines processed the dataset as it was streamed and all resulting candidate events were ingested by GraceDB in the same way astrophysical candidate events are processed in real time. We, therefore, use this MDC dataset and the corresponding entries by the pipelines into GraceDB as a controlled experiement for the signal class.
Moreover, this set up also allows our algorithm to learn out of it how different search pipelines overlap, or not, as they process entire populations of simulated sources. The reader is referred to~\cite{MDCPAPER} for a full description of this MDC dataset.

We will here summarize some key aspects that are relevant for our analysis. Interferometric data acquired by LIGO and Virgo over a 5-week period during their O3 run (January 5, 2020 – February 14, 2020) were used, onto which CBC populations were added via software injections. These injections covered the nominal signal parameter phase space for component masses (1–300 M$_\odot$) and spins. The distance (redshift) distributions of the injections were chosen to be uniform in co-moving volume, with maximum redshifts that varied depending on the source class (BNS, NSBH, or BBH).
Full details of the injection distributions and their astrophysical motivation are provided in Section 3 of the MDC analytics paper~\cite{MDCPAPER}. The event rate of such injections was set to be sufficiently high to allow systematic effects to be measured well beyond statistical uncertainties.
A total of 50,000 signals were generated and
5750 of them were detected by at least one of the four CBC pipelines and with an FAR $\leq 1/\mathrm{hour}$.
From this point of view, the event rate for injections was unphysical. The vast majority of simulated events failed to be detected by the search pipelines primarily because their luminosity distance was beyond the instruments' horizon, thus resulting to low SNR signals on the detectors (and consequently becoming indistinguishable from noise).

\begin{table*}[htb]
    \begin{tabular}{cc}
    \hline
    {Features} & {Description} \\
    \hline\hline
    SNR & Signal-to-Noise Ratio \\
    FAR & False Alarm Rate \\
    Mchirp & Chirp mass of the compact binary coalescence \\
    IFOs SNR Ratio & Ratio of the smallest to the highest SNR among the H1 and L1 \\
    Pipeline Trigger & A binary indicator denoting whether the pipeline produced a trigger for a given event \\
    \hline
    \end{tabular}
    \caption{Summary of the five features used by each one of the four pipelines to collectively form the 20-Dimensional input vectors discussed in Section~\ref{sec:data}.}
    \label{features}
\end{table*}

All simulated events picked up by at least one search pipeline were cross-matched to the injected signal parameters. As part of our pre-processing and analysis, for each one of these events
a 20-dimensional vector (four search pipelines, each providing five features) was formed corresponding to the quantities summarized in Table~\ref{features}. We note that the absence of a trigger by any of the four pipelines was captured in the features vector via zero padding as it carries significance piece of information in combining the pipelines.
Our noise dataset comprises 9,951 events identified by any of the four pipelines considered here (GstLAL, MBTA, pyCBC and cWB-BBH) during real-time running of LVK's O3 run with a FAR greater than or equal to $3.9 \times 10^{-7} \text{ Hz}$ (one per month).
These events were uploaded onto the event database GraceDB but were not confirmed as astrophysical signals or retracted by human vetoing. While there is no absolute source of truth that these events are all the result of noise transients, based on the current rate of astrophysical events, the contamination of astrophysical events present in our noise sample is less that 1\%.

\section{The Best-Of-All-Worlds (\texttt{BOAW}) pipeline} \label{sec:boaw}

The \texttt{BOAW} pipeline introduces an ensemble-based neural network classifier that integrates outputs from multiple CBC search pipelines. By leveraging diverse neural network architectures, the ensemble maximizes discrimination between GW signals and noise. This approach enhances robustness, increases recall, and reduces the false alarm rate (FAR) in real-time CBC detection.\\

\begin{figure}[H]
    \centering
    \includegraphics[scale=0.4]{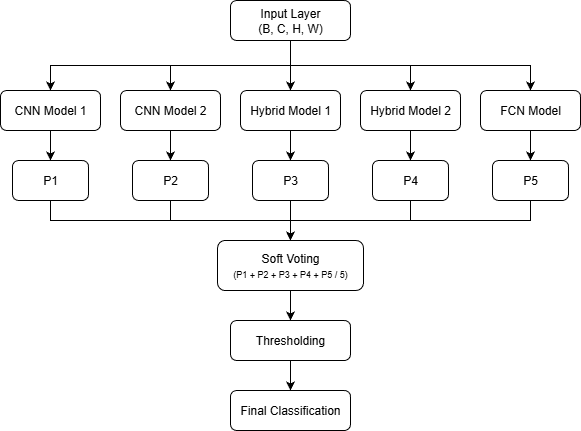}
    \caption{High-level flow diagram of the \texttt{BOAW} pipeline: input features are processed by an ensemble of neural networks and aggregated into a final classification output. The input layer (B,C,H,W) denotes batch size, channels, height, and width, respectively. Each model produces a probability output (P1–P5), which are averaged and thresholded to yield the final classification.}
    \label{fig:NN_model_architecture_level_0}
\end{figure}

To illustrate the overall classification process, Figure~\ref{fig:NN_model_architecture_level_0} provides a high-level flow diagram, demonstrating how input features are processed through each individual model to produce the final classification output.

\subsection{Neural Network Architectures}

The ensemble comprises five distinct neural network models, selected through a Neural Architecture Search (NAS). Three model families form its foundation: the Fully Connected Network (FCN), which uses linear layers to capture global correlations across input features; the Convolutional Neural Network (CNN), which employs 2D convolutional layers to learn localized interactions among features and pipelines; and the Hybrid CNN+FCN, which combines convolutional and fully connected layers to balance local and global pattern recognition.

To accommodate the different architectures, the input vector is reshaped appropriately for each model type. Specifically, for the Fully Connected Network the input is flattened into a one-dimensional 20-element vector. For the Convolutional Neural Network and Hybrid models the input is reshaped into ``spatial" layouts to handle multi-dimensional matrices, with the following configurations: a single-channel $4 \times 5$ 2D matrix represented as (1, 4, 5); four-channel $1 \times 5$ 1D arrays represented as (4, 1, 5); and four-channel $5 \times 1$ 1D arrays represented as (4, 5, 1).

The CNN layers employ a kernel size of 3, stride of 1, and zero padding of 1 to maintain spatial dimensions across layers. Convolutional filters increase in number in deeper layers, allowing the model to learn progressively more complex and abstract features. In contrast, FCN layers reduce the number of neurons per layer, facilitating dimensionality reduction, improved generalization, and the extraction of essential features.

\subsection{Classification Outputs}

Each model supports two output configurations. The first is a single output neuron, which produces a raw logit passed through a sigmoid activation function, yielding a continuous confidence score between 0 and 1. This represents the likelihood of the event being a GW signal.

The second configuration uses {two output neurons}, which produce two logits passed through a softmax activation function, yielding explicit probabilities for the noise and GW-signal classes. These probabilities sum to 1, and the dominant class is determined by the neuron with the highest probability.

The motivation for supporting both configurations is twofold. First, the single-output sigmoid model implements a binary classifier that implicitly treats “signal vs. noise” as a one-dimensional decision boundary; this is often sufficient and computationally efficient. However, it does not explicitly model the noise class. In contrast, the two-output softmax model treats noise and signal as distinct categories and learns separate representations for each. This can be advantageous when the noise class exhibits complex structure or when calibrated class probabilities are required.

Regarding their interpretation, while the sigmoid output may be loosely compared to the “signal” probability from the softmax model, they are not guaranteed to be identical. The sigmoid output reflects the model’s score for the positive class alone, whereas the softmax output reflects a normalized comparison between both classes. Thus, the two configurations can behave differently in practice, and part of our study was to evaluate their relative performance in the context of transient classification.

\subsection{Ensemble Strategy}

The final ensemble model comprises the top five individually trained architectures, with each model contributing to the aggregated prediction to enhance robustness against individual model uncertainties and biases. The ensemble employs a soft voting mechanism, calculating the mean prediction score across all five models. However, the flexibility of the framework allows for the implementation of various decision-making strategies, including soft, weighted, or hard (majority) voting, depending on the models’ confidence scores. This consensus-driven approach not only improves classification performance but also facilitates the quantification of inter-observer variability, offering valuable insights into ambiguous cases and enabling greater control over the final decision.

\subsection{Threshold Selection}

Building upon this framework, we then perform a classification threshold estimation and an uncertainty calculation on the validation set. First, we generate a receiver operating characteristic (ROC) curve and compute the true positive rate (TPR) and false positive rate (FPR) across all potential classification thresholds. The \textit{optimal} threshold is selected using the Euclidean distance to the top-left corner of the ROC space, minimizing

\begin{equation}
\sqrt{(1 - \mathrm{TPR})^2 + \mathrm{FPR}^2}
\end{equation}

From this optimal point, two additional operating thresholds are derived: a \textit{lenient} threshold, obtained by lowering the decision boundary until the true positive rate (TPR) exceeds a predefined target, favoring sensitivity, and a \textit{strict} threshold, obtained by raising the decision boundary until the false positive rate (FPR) drops below a selected value, favoring specificity. In addition to the various threshold options, the selection methodology supports other performance maximization criteria such as 
Youden’s \(J\) statistic,
\begin{equation}
J = \mathrm{TPR} - \mathrm{FPR}
\end{equation}
This approach provides a principled way to balance detection performance across different operational requirements while reporting consistent threshold metrics.

\subsection{Uncertainty Estimation}

Next, we assess model uncertainty by recording each model’s predicted score (ranging from 0 to 1) for every superevent, resulting in a five-element score vector. For each superevent, we compute two measures of ensemble disagreement: (i) the standard deviation (STD) of the scores, and (ii) the score range (maximum minus minimum score). Across all superevents, we then calculate the grand mean and standard deviation of each measure, denoted as $\mathrm{mean}_{\mathrm{std}}$, $\mathrm{std}_{\mathrm{std}}$, $\mathrm{mean}_{\mathrm{range}}$, and $\mathrm{std}_{\mathrm{range}}$. The uncertainty threshold is defined as
\begin{equation}
\mathrm{threshold} = \mathrm{mean} + k \cdot \mathrm{std}
\end{equation}
where $k = 1.96$. This value corresponds to the 95\% coverage interval of a normal distribution, meaning that--under the assumption that ensemble-disagreement measures are approximately Gaussian--about 95\% of typical (low-uncertainty) superevents are expected to fall below this threshold.
Superevents with disagreement values above this cutoff therefore lie in the upper 5\% tail of the distribution and are flagged as unusually uncertain.By default, we apply this threshold using the score range, though the same formulation can be applied to the standard deviation of scores. Superevents exceeding the threshold are marked as having elevated ensemble disagreement, signaling increased predictive uncertainty.

Harnessing the diverse pattern-recognition capabilities of each model architecture, the ensemble classifier achieves reliable and accurate performance in distinguishing GW signals from noise, characterized by high recall and a low false-alarm rate.

\section{Training}
\label{sec:training}

The proposed network architecture was trained on a dataset in which the positive class comprises MDC injections, while the negative class is composed exclusively of noise samples drawn from the O3 dataset. In total, the training dataset includes 8,951 noise samples and 4,750 positive samples as described in Section ~\ref{sec:data}, leaving out 1000 samples from each class in order to form a test set for subsequent evaluation.

\subsection{Preprocessing}

Our analysis started with a preprocessing phase designed to ensure data quality, consistency, and suitability for downstream ML tasks. This phase consisted of several steps. {Event filtering} was applied to remove events missing critical data fields, including the Superevent identifier, SNR, FAR, and chirp mass. {Feature engineering} was subsequently performed to construct derived features such as (i) the ratio of minimum-to-maximum SNR across detectors, (ii) the inverse false alarm rate, i.e., \text{iFAR} = 1/\text{FAR},
and (iii) a pipeline trigger flag, a binary indicator of whether a pipeline produced a trigger. {Feature scaling} was then applied using min–max normalization to map all features into the [0,1] range, with domain-specific bounds listed in Table~\ref{tab:feature_scaling_bounds}. Finally, {class balancing} was addressed through stratified sampling to ensure that training and validation splits preserved the signal–noise proportions; additional steps to manage class imbalance during training are described in Section~\ref{subsec:training_protocol}.

\begin{table}[ht]
\centering
\begin{tabular}{ccc}
\toprule
\hline
{Feature} & {Lower Bound} & {Upper Bound} \\ 
\hline\hline
SNR             & 1.0                  & 100.0         \\
FAR (Hz)        & $3.16 \times 10^{-10}$ & $2.78 \times 10^{-4}$ \\
$M_{\text{chirp}}$ ($M_\odot$) & 0.0 & 200.0 \\
\hline
\toprule
\end{tabular}
\caption{Bounds used for min--max scaling of SNR, FAR, and chirp mass. For $\text{iFAR}$, the reciprocal of these FAR bounds is applied.}
\label{tab:feature_scaling_bounds}
\end{table}

\subsection{Training Protocol}
\label{subsec:training_protocol}

In our training procedure, we first employed a 5-fold stratified cross-validation setup to guide both model architecture selection and hyperparameter tuning, while also ensuring that each fold adequately represented the class distribution. Each fold was then further partitioned into an 80/20 split for training and validation.

Because the noise population is much larger than the signal population, individual cross-validation folds necessarily contain only a subset of the available noise examples. After selecting the optimal model and hyperparameters, we proceeded to a final training stage in which the entire noise dataset was incorporated through dynamic resampling: at each training epoch, a different subset of noise examples was drawn from the full noise pool. This approach exposes the model to a broad and diverse set of noise morphologies throughout training while maintaining a fixed, balanced ratio of signal and noise within each batch.

\subsection{Loss Function}

To enhance sensitivity to GW signals, the loss function used during training is a scaled Binary Cross-Entropy (BCE), weighted by the mean inverse False Alarm Rate (iFAR) of the superevent. This dynamic scaling penalizes misclassification of significant events, thereby improving the model's sensitivity to real events.

The Binary Cross-Entropy (BCE) loss for a single prediction \( i \) is given by:
\begin{equation}
\text{BCE}(p_i, y_i) = - \left( y_i \log(p_i) + (1 - y_i) \log(1 - p_i) \right)
\end{equation}
where \( p_i \) denotes the predicted probability for the \( i \)-th sample and \( y_i \) denotes the true label for the \( i \)-th sample.

Let $\text{iFAR}_{\text{batch}}$ denote the mean inverse False Alarm Rate (iFAR) for a batch, calculated as
\begin{equation}
\text{iFAR}_{\text{batch}} = 
\frac{\sum_{j=1}^{M} \text{iFAR}_j}{\sum_{j=1}^{M} \mathbf{1}(\text{iFAR}_j \neq 0)},
\end{equation}
where M is the number of individual events in the batch (number of superevents in the batch multiplied by the number of events in a superevent, in this case 4) $\text{iFAR}_j$ is the inverse false alarm rate value for the $j$-th event in the batch, and $\mathbf{1}(\text{iFAR}_j \neq 0)$ is the indicator function that equals 1 if $\text{iFAR}_j \neq 0$, ensuring that only non-zero (non-padded) events are included in the denominator.

The final loss function is scaled by this mean iFAR and computed as
\begin{equation}
\mathcal{L} = \frac{1}{N} \sum_{i=1}^{N} \text{BCE}(p_i, y_i) \cdot \left(1 + \lambda \cdot \text{iFAR}_{\text{batch}} \right),
\end{equation}
where \( N \) is the number of superevents in the batch and \( \lambda \), the iFAR modulation coefficient, is a hyperparameter that determines the level of scaling applied to the loss based on iFAR.

\subsection{Hyperparameter Optimization}

We employed a systematic hyperparameter search strategy, introducing one incremental change at a time to isolate its impact on performance. At each stage, we evaluated both training and validation metrics to ensure any modification led to meaningful gains; when certain optimizations-such as bottleneck architectures, random weight initializations, specific optimizers (e.g., SGD), gradient clipping, or batch normalization, showed limited benefit, we reverted to simpler setups.

After converging on a well-defined set of hyperparameter search space (listed in Table~\ref{tab:hyperparams}), we conducted a randomized search procedure, examining multiple neural network architectures by training each candidate for 200 epochs across five folds.

\begin{figure}
    \centering
    \includegraphics[scale=0.5]{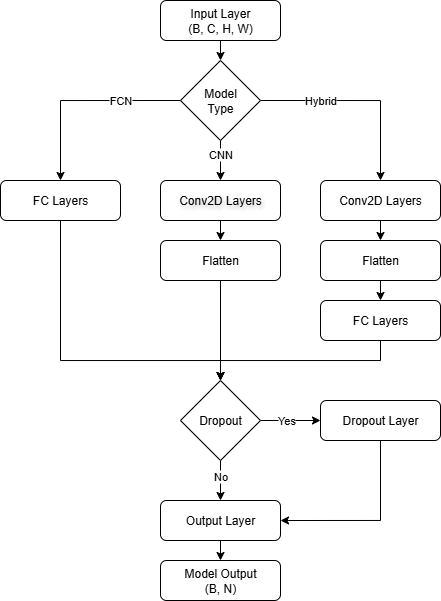}
    \caption{Training flowchart of the architecture framework for systematic hyperparameter optimization, featuring conditional branching to select between FCN, CNN, or Hybrid models, with configurable dropout and uniform input–output handling across all variants. The input layer (B,C,H,W) denotes batch size, channels, height, and width, while the model output (B,N) represents batch size and the number of output classes.}
    \label{fig:FlexibleNN_NN_architecture_diagram}
\end{figure}

\subsection{Final Model Selection}
To select models for the ensemble formation, we first established a performance baseline by identifying the top 20 models within each subtype, i.e. FCN, CNN, and Hybrid, ranked by Youden’s statistic. These models exhibited consistently high and comparable performance, making raw metrics alone insufficient for prioritization. From each subtype, the best-performing model was selected, after which the remaining models in that category were iteratively compared against it to find the most diverse partner. Diversity was quantified through an automated scoring procedure that combined multiple factors: architectural differences (input/output shapes, convolutional and fully connected layer counts), training configuration variations (activation functions, optimizers, initializers, batch sizes), and data processing differences (output types, label inversions). Each factor contributed a weighted score, with architectural variations emphasized to ensure the ensemble captured complementary learning patterns rather than redundant predictions. The final ensemble comprised two CNNs, two Hybrid models, and a single FCN, with the latter limited to one instance since CNN and Hybrid subtypes consistently delivered stronger performance. By combining models across all three subtypes, the ensemble achieved greater robustness, ultimately reducing the risk of correlated failure modes.

\begin{figure}
    \centering
    \includegraphics[width=\columnwidth, keepaspectratio]
    {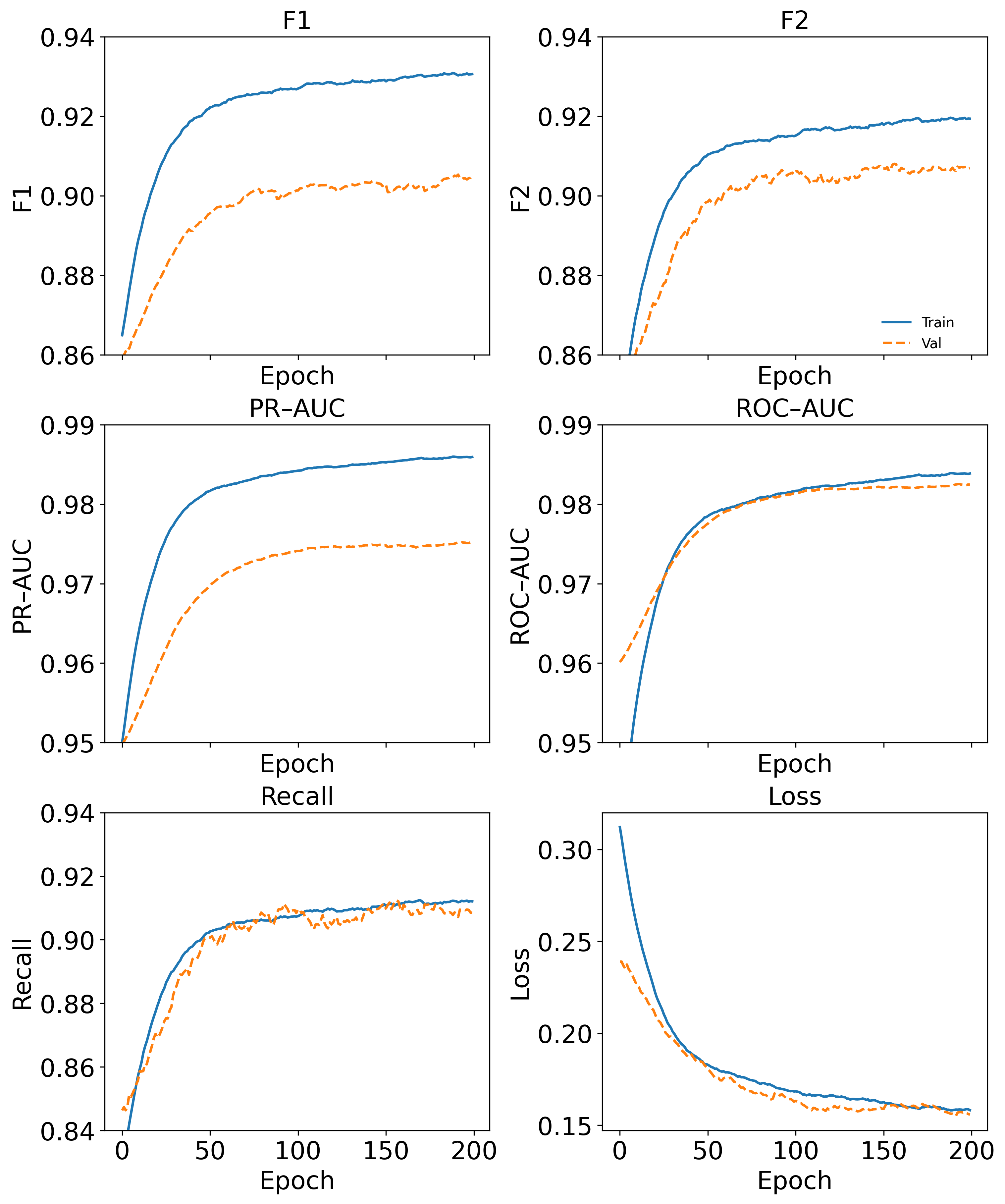}
    \caption{Training and validation metrics for one representative model in the ensemble, showing the evolution of key performance metrics across training epochs.}
    \label{fig:training_metrics}
\end{figure}

Shortlisted models were retrained on an expanded split (90\% training, 10\% validation) for 200 epochs. Model checkpoints were saved when validation metrics improved, subject to a safeguard requiring training performance to remain at least as high, preventing overfitting to favorable validation subsets. Metrics monitored included loss, recall, F1, F2, ROC–AUC, and PR–AUC, together with additional diagnostic metrics used to track training stability, with representative learning curves shown in Fig.~\ref{fig:training_metrics}. Based on these evaluations, the five models with the highest ROC-AUC--prioritizing configurations that maximized TPR while minimizing FPR, balanced with diversity, were selected as the ensemble members.

\makeatletter
\def\heavyrulewidth{0.08em}
\def\lightrulewidth{0.05em}
\def\cmidrulewidth{0.03em}
\makeatother

\begin{table*}[htb]
\centering
\small
\begin{tabular}{llll}
\hline
{Parameter} & {Description} & {Prior} & {Limits} \\
\hline\hline
Learning Rate      & Learning rate for optimizer                      & Uniform      & $(5\times10^{-5}, 10^{-2})$ \\
Batch Size         & Mini-batch size                                  & Uniform      & $(32, 512)$ \\
Model Type         & Type of model architecture                       & Categorical  & \{\texttt{CNN}, \texttt{FCN}, \texttt{Hybrid}\} \\
Input Shape        & Input shape handling                             & Categorical  & \{(1,4,5), (4,1,5), (4,5,1), (20)\} \\
CNN Layers         & Number of convolutional layers                   & Uniform      & $(1, 5)$ \\
CNN Filters        & Channel configuration per conv layer             & Uniform      & $(16, 1024)$ \\
FCN Layers         & Number of fully connected layers                 & Uniform      & $(1, 3)$ \\
FCN Neurons        & Neuron count(s) in FC layers                     & Uniform      & $(2, 128)$ \\
Dropout Layer      & Whether to include dropout                       & Boolean      & \{\texttt{True}, \texttt{False}\} \\
Dropout Rate       & Dropout fraction when enabled                    & Uniform      & $[0.3, 0.5]$ \\
Activation Function& Activation function selection                    & Categorical  & \{\texttt{relu}, \texttt{l\_relu}, \texttt{prelu}, \texttt{gelu}\} \\
Optimizer          & Optimization algorithm                           & Categorical  & \{\texttt{adam}, \texttt{adamW}\} \\
Scheduler          & Learning rate scheduling strategy                & Categorical  & \{\texttt{reduce\_on\_plateau}, \texttt{cosine}, \texttt{none}\} \\
Initializer        & Weight initialization method                     & Categorical  & \{\texttt{he\_norm}, \texttt{he\_unif}, \texttt{xavier\_norm}, \texttt{xavier\_unif}\} \\
\makecell[tl]{\textit{iFAR}\\scaling factor} & Scaling factor for inverse FAR (iFAR) & Uniform & $[0.0, 5.0]$ \\
Output Size        & Output layer dimension                           & Categorical  & \{1, 2\} \\
Invert Labels Flag & Train with inverted labels                       & Boolean      & \{\texttt{True}, \texttt{False}\} \\
\hline
\end{tabular}
\caption{Hyperparameter ranges and choices used in our random-search process.}
\label{tab:hyperparams}
\end{table*}

\section{Results} \label{sec:results}

We evaluated the performance of the ensemble neural network on a held-out test dataset comprising 1000 MDC injections (positive class) and 1000 O3 noise triggers (negative class). Event classification used the optimal decision threshold of 0.38 (see Section~\ref{sec:boaw}), corresponding to classifying scores $\geq 0.38$ as signals (injections). Model outputs range from 0 to 1, providing a continuous measure of confidence. Evaluation metrics include ROC curves, confusion matrices, score distributions, and correlation analyses with key input features (SNR, chirp mass, FAR). In particular, we note that the current operating procedure within LVK for sending GW alerts involves a classification thresholding on FAR~\cite{USERGUIDE}. We therefore provide comparisons against FAR-based classification.

\subsection{ROC Performance Against Pipelines}

\begin{figure}[htbp]
  \centering
  \includegraphics[width=\linewidth]{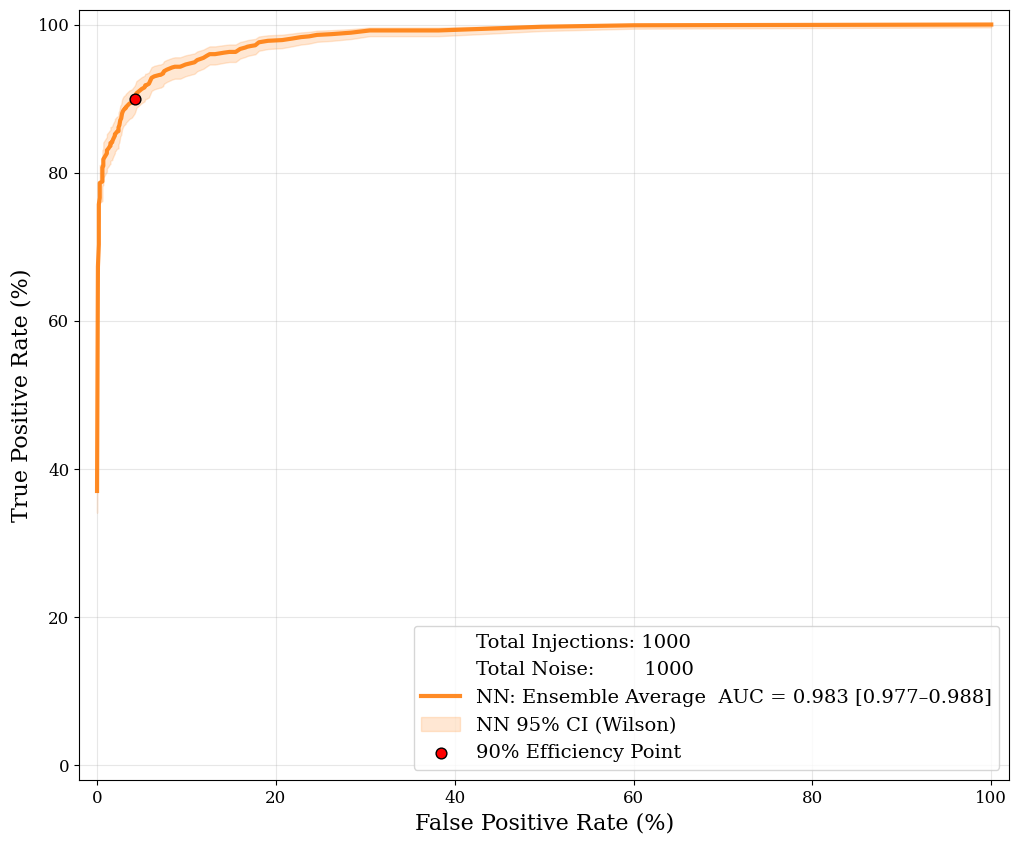}
  \caption{ROC curve of the neural network ensemble evaluated on the full test dataset (1000 injections and 1000 noise events). The shaded region represents 1-$sigma$ errors.
  The ensemble achieves an AUC of 0.983 with confidence interval [0.977–0.988].}
  \label{fig:roc_nn_full_test}
\end{figure}

We begin by presenting the standalone performance of the neural network ensemble on the full test dataset, shown in Fig.~\ref{fig:roc_nn_full_test}. This ROC curve summarizes the classifier behavior across all decision thresholds, illustrating the trade-off between true positive rate (TPR) and false positive rate (FPR). The shaded region represents 1-$\sigma$ errors.
The ensemble achieves an AUC of 0.983, with a confidence interval of [0.977–0.988], demonstrating good discrimination between injections and noise on the full test set.

\begin{figure*}[htbp]
  \centering
  \begin{subfigure}[t]{0.48\textwidth}
    \centering
    \includegraphics[width=\linewidth]{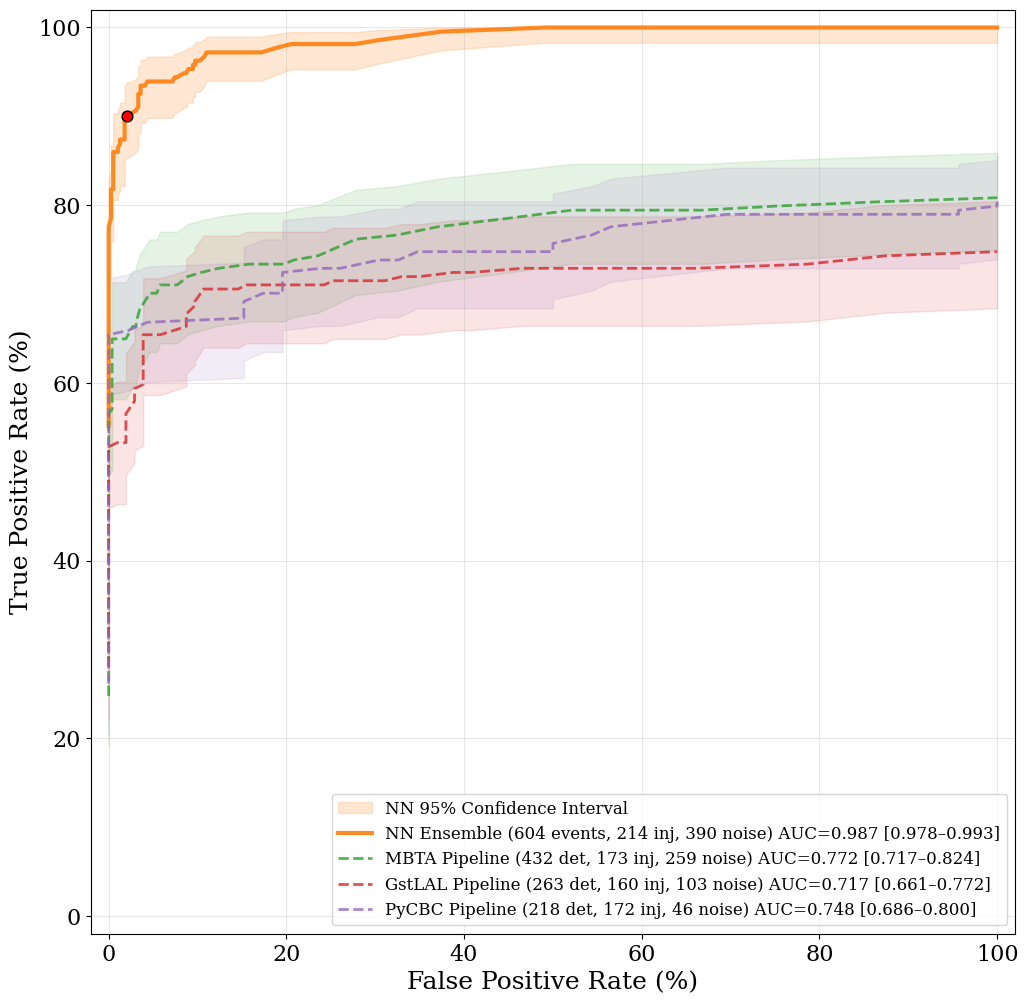}
    \caption{Performance for superevents with chirp mass below $1.6~M_\odot$; no cWB-BBH triggers are considered in this analysis.}
    \label{fig:roc_nn_vs_pipelines_below_cwb}
  \end{subfigure}
  \hfill
  \begin{subfigure}[t]{0.48\textwidth}
    \centering
    \includegraphics[width=\linewidth]{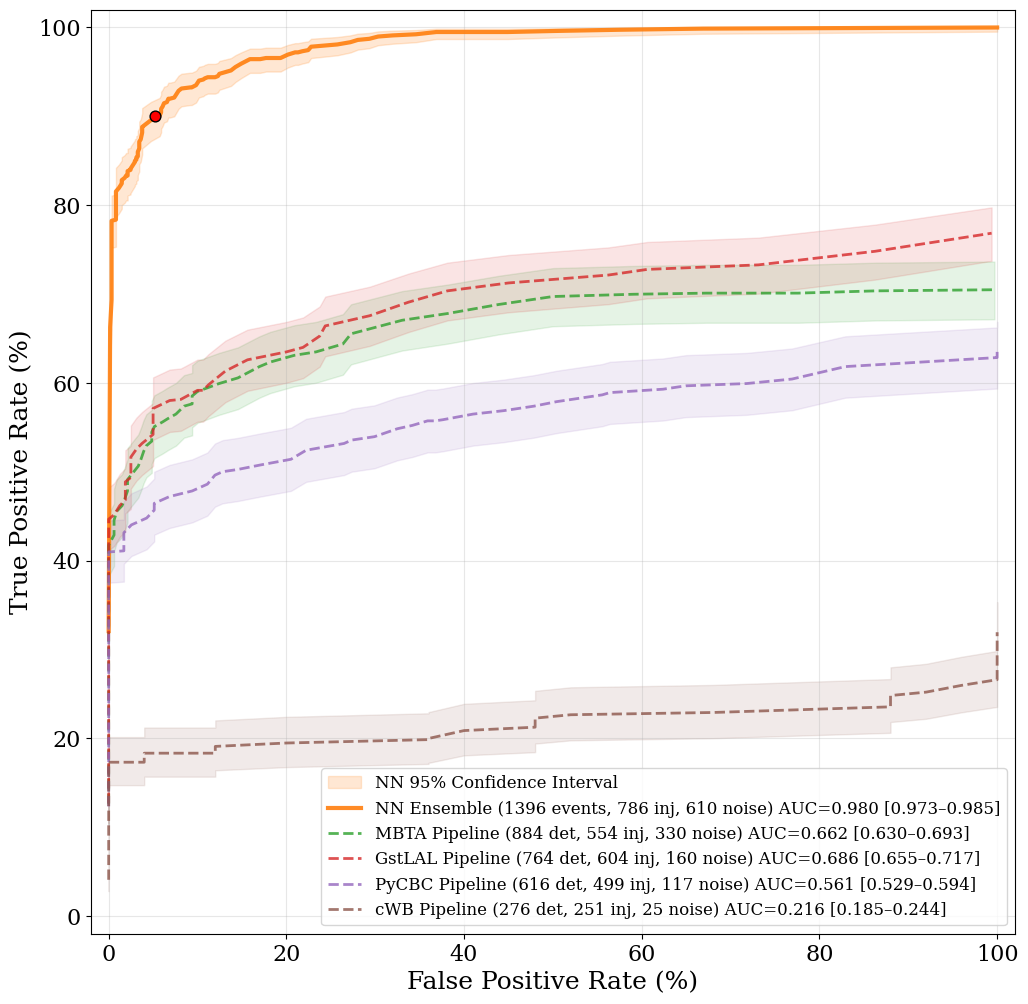}
    \caption{Performance for superevents with chirp mass at or above $1.6~M_\odot$. All four CBC methods are considered in this analysis.}
    \label{fig:roc_nn_vs_pipelines_above_cwb}
  \end{subfigure}

  \caption{ROC comparison of the neural network ensemble (orange) with individual search pipelines. Across both chirp mass ranges, the ensemble provides superior discrimination, achieving higher true positive rates at substantially lower false positive rates than any individual pipeline. The red dot indicates the point corresponding to 90\% true positive rate/efficiency for reference.}
  \label{fig:roc_nn_vs_pipelines}
\end{figure*}

Figure~\ref{fig:roc_nn_vs_pipelines} compares the neural network ensemble classifier with the individual search pipelines (GstLAL, MBTA, PyCBC, and cWB-BBH), with results shown separately for superevents corresponding to chirp mass below and above $1.6~M_\odot$.
These two regimes of chirp mass values were established empirically and were driven by the fact the cWB-BBH method is primarily targeting BBH systems.
The exact value of chirp mass was determined by the signal injection with the lowest chirp mass value that resulted to a trigger by the cWB-BBH method, although the results of this analysis are not critically dependent on this.
To ensure a fair comparison between methods that evaluate different subsets of events, we adopt denominators tailored to this setting. For the TPR, the denominator is the total number of astrophysical injections in the corresponding chirp mass subset. This penalizes pipelines for missed detections and is the reason their ROC curves do not reach the upper-right corner: pipelines do not trigger on all injected astrophysical signals, so their maximum achievable TPR is inherently below 100\%. For the FPR, the denominator is the number of noise events detected by that pipeline only, ensuring that a pipeline is not penalized for noise triggers produced by other algorithms. These choices isolate each method’s intrinsic behavior: pipelines are evaluated strictly on the events they identify, while the ensemble is evaluated on all superevents.

In the low chirp mass regime (Fig.~\ref{fig:roc_nn_vs_pipelines_below_cwb}), where only GstLAL, MBTA, and PyCBC are sensitive,
the ensemble ROC curve lies uniformly above all single-pipeline curves. At the 90\% TPR operating point, the ensemble achieves a FPR of 2\%, supported by a high AUC of 0.99, demonstrating substantially stronger discrimination than any individual pipeline. For comparison, the individual pipelines achieve considerably lower AUC values that range from 0.72 to 0.77, highlighting the ensemble’s clear signal vs noise discrimination power in the low chirp mass regime. 
In the high chirp mass regime (Fig.~\ref{fig:roc_nn_vs_pipelines_above_cwb}), where cWB-BBH becomes sensitive, the ensemble again provides the strongest performance (AUC = 0.98, 5\% FPR at 90\% TPR), while several individual pipelines show limited discrimination ability.

\begin{figure*}[htbp]
  \centering
  \begin{subfigure}[t]{0.48\textwidth}
    \centering
    \includegraphics[width=\linewidth]{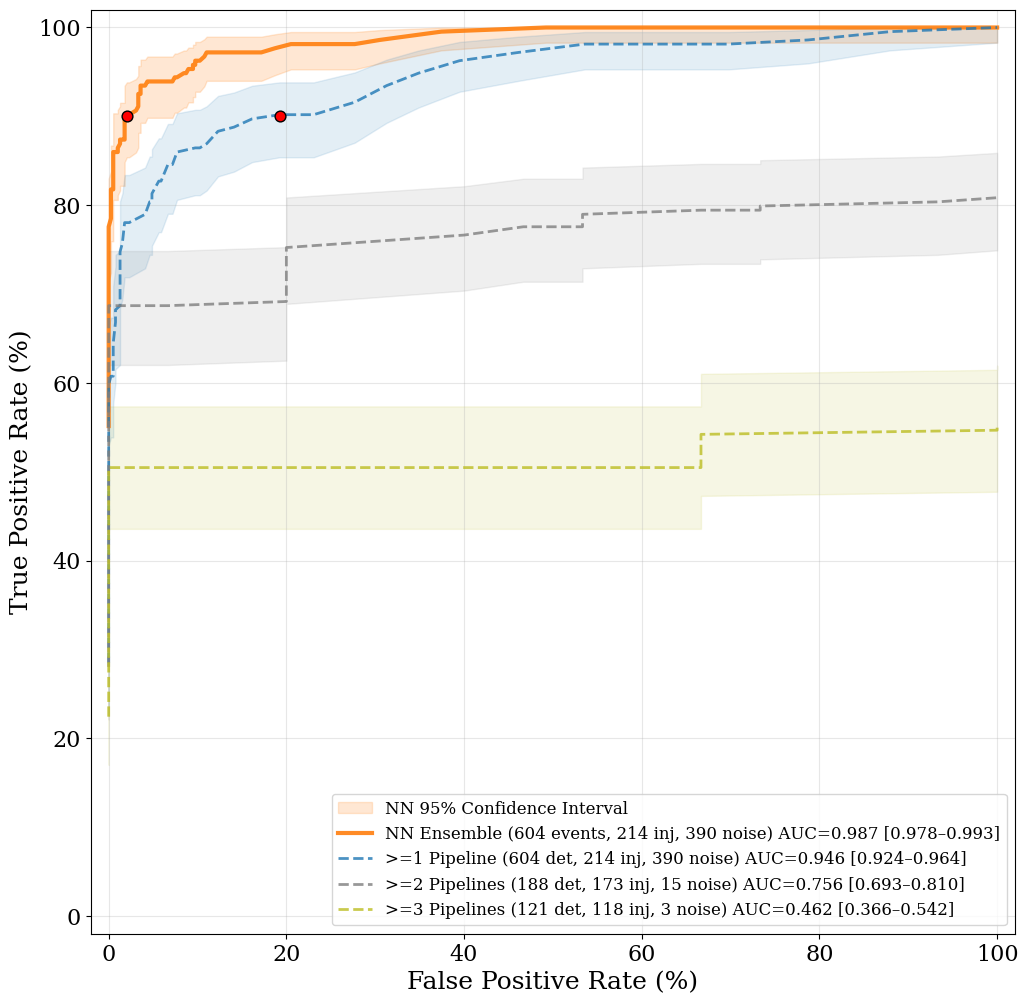}
    \caption{Performance for superevents with chirp mass below $1.6~M_\odot$; no cWB-BBH triggers are considered in this analysis.}
    \label{fig:roc_nn_vs_logical_below_cwb}
  \end{subfigure}
  \hfill
  \begin{subfigure}[t]{0.48\textwidth}
    \centering
    \includegraphics[width=\linewidth]{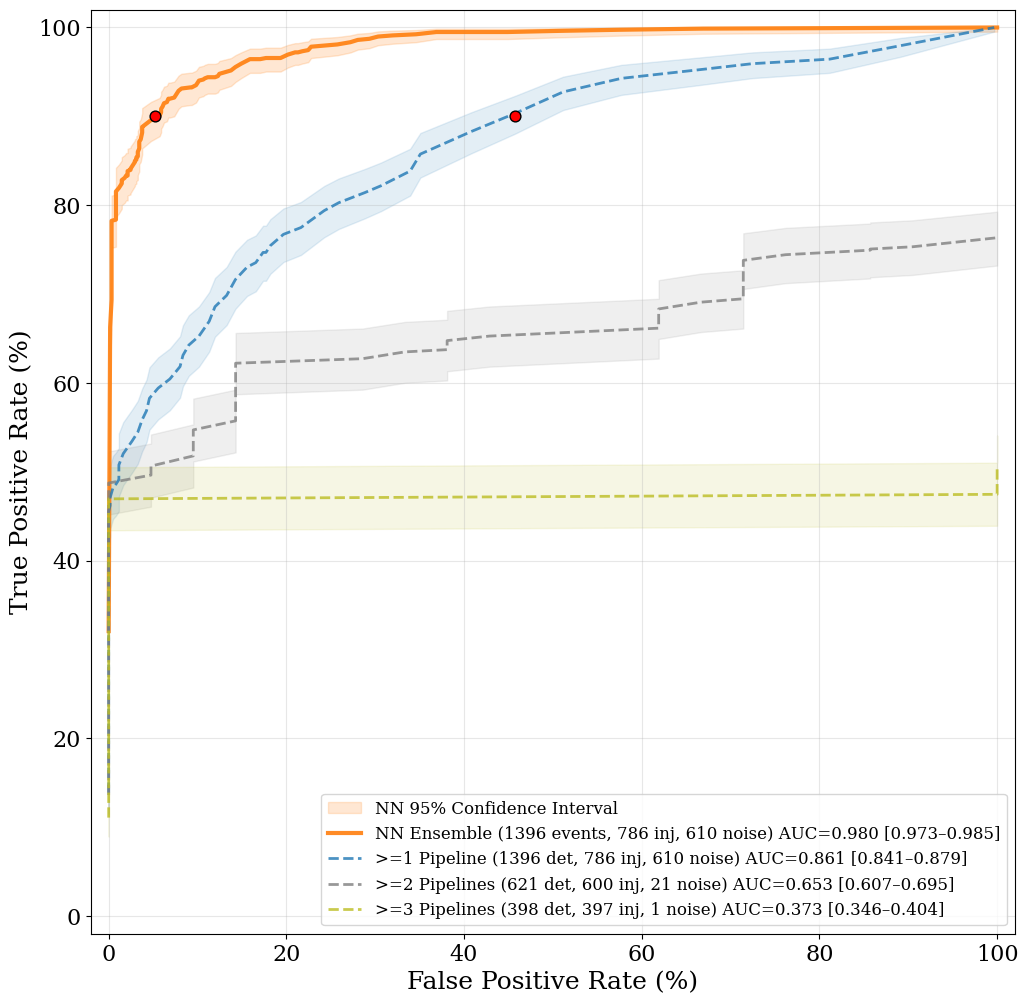}
    \caption{Performance for superevents with chirp mass at or above $1.6~M_\odot$. All four CBC methods are considered in this analysis.}
    \label{fig:roc_nn_vs_logical_above_cwb}
  \end{subfigure}

  \caption{ROC comparison of the ensemble (orange) with logical pipeline combinations. Across both chirp mass regimes, the ensemble yields substantially lower false positive rates, including nearly an order-of-magnitude improvement near 90\% true positive rate/efficiency.}
  \label{fig:roc_nn_vs_logical_combinations}
\end{figure*}

Figures~\ref{fig:roc_nn_vs_logical_below_cwb} and \ref{fig:roc_nn_vs_logical_above_cwb} compare the neural network ensemble with logical combinations of pipelines based on FAR classifications, again separating superevents by chirp mass. In both chirp mass regimes, the ensemble consistently outperforms the coincidence rules: its ROC curve remains above the $\geq 1$, $\geq 2$, and $\geq 3$ pipeline combinations across operating points, with particularly large gains near the 90\% TPR operating point. For the low chirp mass subset, the ensemble achieves a FPR of only $\sim$2\%, whereas the $\geq 1$ logic yields $\sim$19\%. This gap widens in the high chirp mass regime: the ensemble’s FPR remains below $\sim$5\%, while the pipeline combination exceeds $\sim$46\%, reflecting nearly a tenfold degradation in specificity.

As stricter pipeline agreement requirements are imposed, the number of events contributing to the ROC curves decreases correspondingly, as shown in the figure legends. The most extreme case requiring all four pipelines to trigger, was not plotted. In our test dataset, every event satisfying this $\geq 4$ condition corresponded to an astrophysical injection, and no noise event was simultaneously detected by all pipelines. Under our ROC conventions, the denominator for the FPR (the number of detected noise triggers) is therefore zero, making the FPR undefined. Although this rule would appear to yield perfect specificity, it applies to an extremely small, signal-only subset and provides no meaningful comparison to classifiers that operate on the full set of superevents.

Across both chirp mass regimes, the ensemble provides superior ROC performance without the sensitivity–specificity trade-offs characteristic of logical pipeline combinations. Unlike coincidence rules which operate only on restricted and pipeline-dependent subsets of events, the ensemble evaluates every superevent, yielding more reliable and comprehensive discrimination between astrophysical signals and terrestrial noise.

\subsection{Ensemble Score Distribution}

\begin{figure}[htbp]
\centering
\includegraphics[width=0.99\columnwidth]{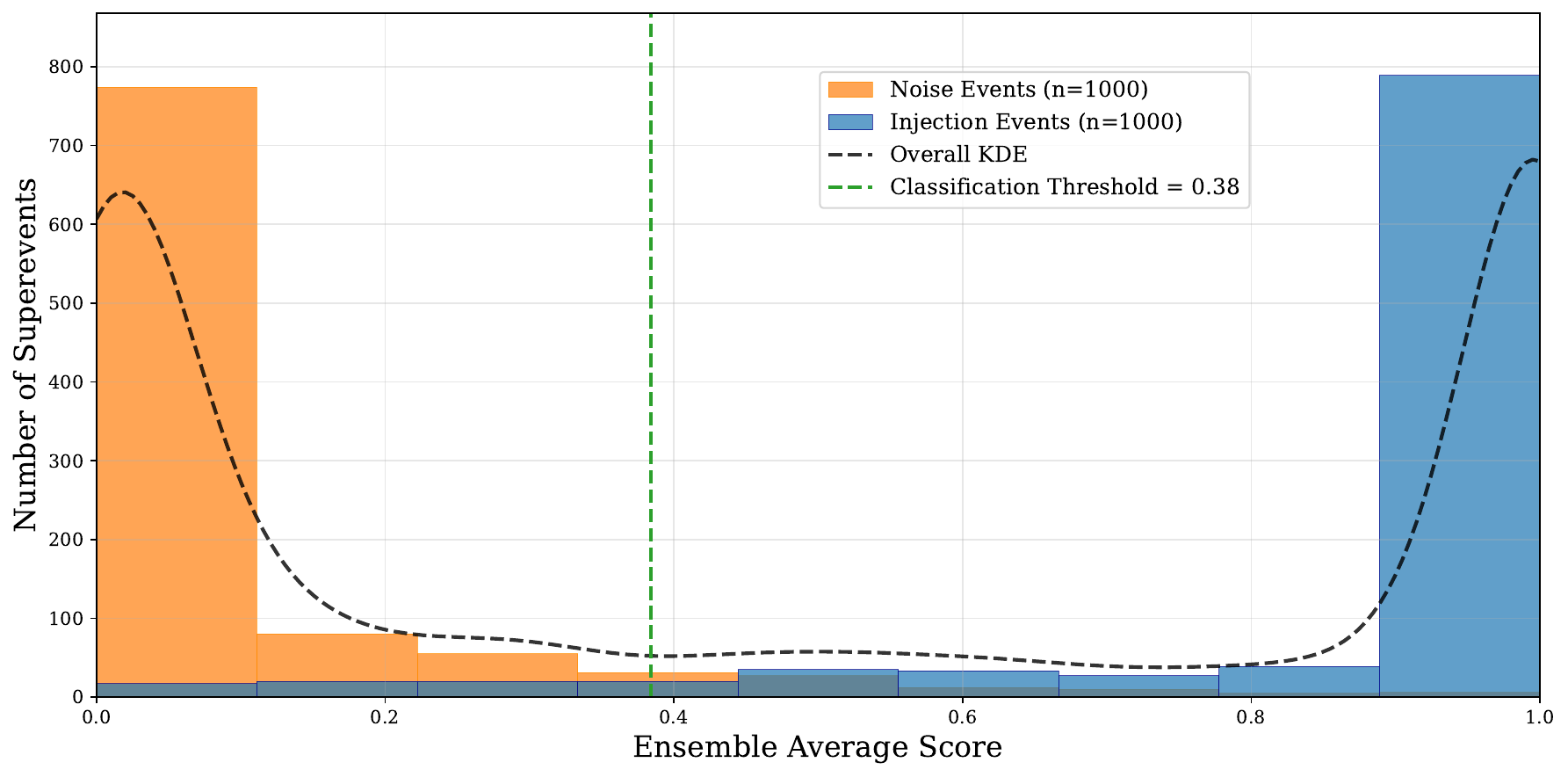}
\caption{Distribution of ensemble scores for noise (orange) and injections (blue). Peaks near 0 and 1 confirm strong separation. The black dashed line represents the overall kernel density estimate (KDE), while the green vertical line at 0.38 marks our nominal threshold.}
\label{fig:ensemble_score_distribution}
\end{figure}

The ensemble score distribution plot in Fig.~\ref{fig:ensemble_score_distribution} shows a clear bimodal distribution of the model scores, with one peak near 0 and another near 1. This bimodality confirms the model's ability to effectively separate noise from real astrophysical events, as scores near 0 correspond to noise and those near 1 to real events. The presence of a sharp transition with very few scores in the middle range further reinforces the model’s confidence in classification. The dashed line at the 0.38 threshold indicates the decision boundary, showing that the vast majority of noise events are classified below this threshold, while most injection events are above it.

\subsection{Confusion Matrices}

\begin{figure*}[htbp]
\centering
\includegraphics[width=0.99\textwidth]{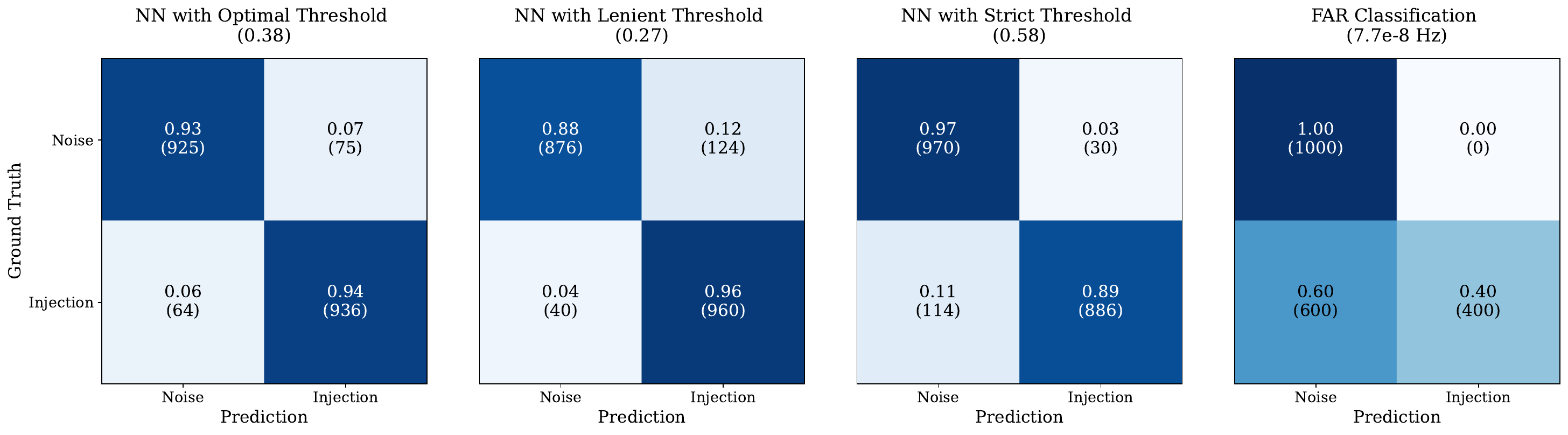}
\caption{Confusion matrices comparing neural network (NN) ensemble performance under different threshold strategies versus FAR-based classification on the test dataset. Neural network thresholds show: optimal (0.38), lenient (0.27), and strict (0.58), while FAR classification uses a threshold of $7.7 \times 10^{-8} \mathrm{Hz}$ (one per five months). Values show classification rates with event counts in parentheses. The positive class of the test set consists of 1000 injected signals recovered by at least one of the four pipelines considered; however, recovery does not require the events to meet the high-significance FAR threshold.}
\label{fig:conf_matrix_O3_test_set}
\end{figure*}

The confusion matrices in Fig.~\ref{fig:conf_matrix_O3_test_set} demonstrate the performance of our neural network ensemble under different threshold strategies compared to FAR-based classification on the test dataset. The optimal threshold approach (threshold=0.38) achieves great overall performance with 93\% specificity and 94\% sensitivity.
The lenient threshold strategy (threshold=0.27) prioritizes sensitivity, achieving 96\% recall at the cost of increased false positives, making it suitable for applications where missing true signals is more costly than processing false alarms. Conversely, the strict threshold (threshold=0.58) maximizes specificity at 97\% while maintaining 89\% sensitivity, making it appropriate for scenarios requiring high confidence in positive detections.

Notably, the FAR-based classification with the public alert threshold used in the MDC replay of the O3 data (that provided the input for this analysis~\cite{USERGUIDE}) of one per five months (FAR $< 7.7 \times 10^{-8} \mathrm{Hz}$) exhibits different behavior, achieving perfect specificity, i.e. zero false positives, but at the cost of dramatically reduced sensitivity, correctly identifying only 400 of 1000 injection events. This contrast highlights the conservative nature of FAR-based thresholds in GW detection, which prioritize extremely low false alarm rates at the expense of detection efficiency. Our neural network ensemble demonstrates superior balanced performance, with even the strict threshold strategy detecting over twice more true signals than the FAR approach while maintaining comparable false positive rates. These results validate the ensemble's ability to provide flexible, application-specific classification strategies that significantly outperform traditional statistical methods in terms of detection sensitivity.

\subsection{True Positive Recovery as a Function of Signal-to-Noise Ratio}
\label{subsec:tp_snr_distribution}

While the confusion matrices in Fig.~\ref{fig:conf_matrix_O3_test_set} provide an aggregate comparison of classification performance, they do not reveal how true positive recoveries are distributed across signal strength. To further characterize the origin of the performance gains achieved by the neural network ensemble, we examine the distribution of correctly classified injection events as a function of network signal-to-noise ratio (SNR).

\begin{figure}[htbp]
\centering
\includegraphics[width=0.99\columnwidth,height=0.75\textheight,keepaspectratio]{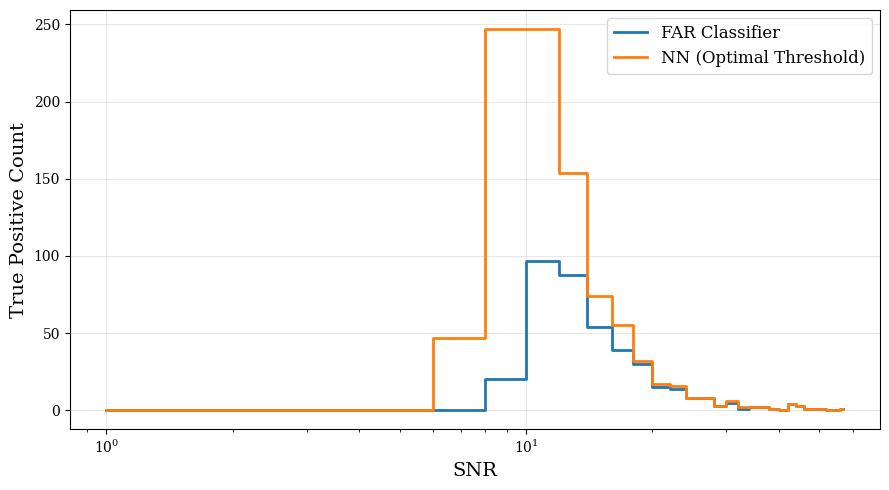}
\caption{Distribution of true positive as a function of network SNR for the neural network ensemble operating at the optimal threshold (0.38) and for the FAR-based classifier ($\mathrm{FAR} < 7.7 \times 10^{-8}\,\mathrm{Hz}$, or one per five months). The ensemble recovers a larger fraction of low-SNR signals, consistent with the increased true positive counts observed in Fig.~\ref{fig:conf_matrix_O3_test_set}.}
\label{fig:tp_snr_distribution}
\end{figure}

Figure~\ref{fig:tp_snr_distribution} shows the true positive (TP) superevent counts binned in SNR for the neural network ensemble operating at the optimal threshold (0.38) and for the FAR-based classifier using a threshold of $7.7 \times 10^{-8}\,\mathrm{Hz}$ (one per five months). Only injected signals that are correctly classified as astrophysical are included.
Our method exceeds the performance of an FAR-based event selection across all SNR bins, markedly increasing the ability to recover lower SNR astrophysical injections.
This behavior is weakly dependent on the exact value of FAR thresholds chosen within
the range of 1 per ten months to 1 per five months.
These values covers most of the values the LVK chose in operating the public astronomical alerts system over the course of recent observing runs and MDC simulation runs they invoked for studying low latency performance of their system~\cite{MDCPAPER,USERGUIDE}

This behavior explains the disparity in true positive counts observed in Fig.~\ref{fig:conf_matrix_O3_test_set}. As shown there, the neural network with optimal threshold correctly identifies 936 of the 1000 injected signals in the test set, compared to 400 recoveries by the FAR classifier. Figure~\ref{fig:tp_snr_distribution} demonstrates that the majority of this improvement arises from enhanced sensitivity in the low-SNR regime, rather than from marginal gains at high SNR where both methods perform comparably.

Importantly, this low-SNR recovery capability is astrophysically relevant: weaker signals correspond to more distant or less favorably oriented sources, and their recovery directly impacts the observable volume and completeness of GW catalogs. By learning a multivariate decision boundary that combines SNR, FAR, chirp mass, and additional pipeline-derived features, the ensemble is able to retain sensitivity to these challenging events without incurring the dramatic loss of specificity characteristic of relaxed FAR thresholds.

Taken together, the confusion-matrix results (Fig.~\ref{fig:conf_matrix_O3_test_set}) and the SNR-resolved TP distribution (Fig.~\ref{fig:tp_snr_distribution}) demonstrate that the neural network ensemble does not merely increase the total number of detected signals, but does so in a physically meaningful way by extending detection efficiency into the low-SNR regime where FAR-based methods are inherently limited.

\subsection{Performance on Astrophysical Categories}

Further breakdowns of the confusion matrix using the optimal threshold by astrophysical categories reveal consistent high performance across all event types, with some notable variations. For BBH events (n=615), the model achieves a classification of 96.3\% sensitivity and 81.3\% specificity. BNS events (n=768) demonstrate the strongest performance with 92.4\% sensitivity and 96.9\% specificity. NSBH events (n=617) show good performance with 90.0\% sensitivity and 91.6\% specificity. Notably, BNS events exhibit the lowest false positive rate (3.1\%) among all categories, while BBH events show the highest false positive rate (18.7\%).

\subsection{Real-Time Suitability}

To test \texttt{BOAW}’s suitability for real-time deployment, we evaluated its sensitivity (TPR) and fall-out (FPR) in comparison to the combination of the individual search pipelines with a FAR classification threshold of $7.7 \times 10^{-8} \mathrm{Hz}$ 
(one per five months) used within a factor of two in LVK’s production environment in recent observing runs and MDC replays~\cite{MDCPAPER,USERGUIDE}.

\begin{figure*}[htbp]
\centering
\includegraphics[width=0.99\textwidth]{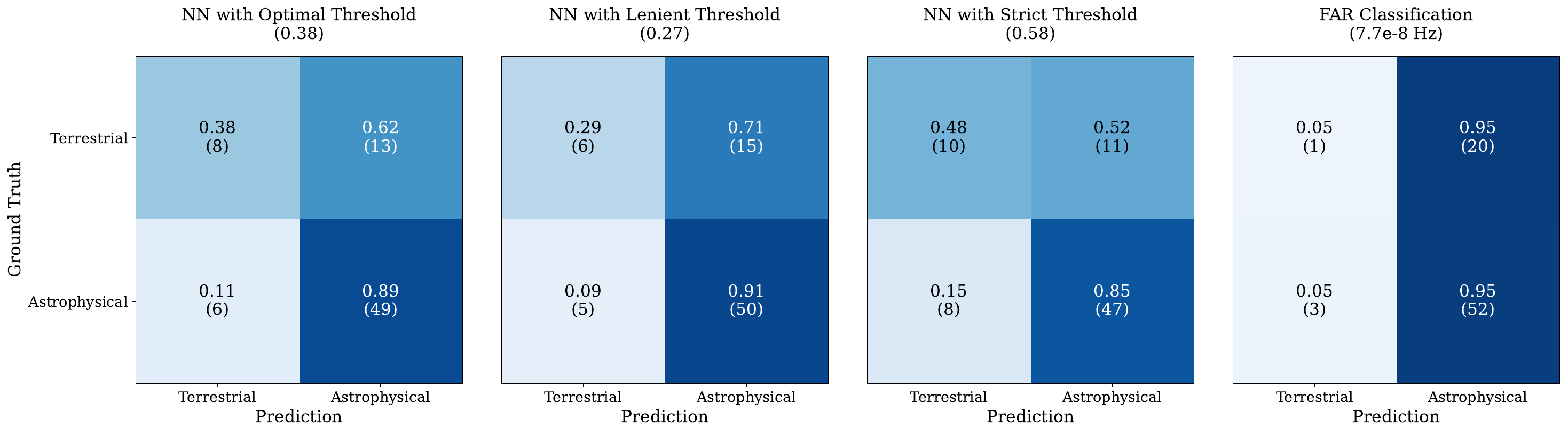}
\caption{Confusion matrices comparing neural-network (NN) ensemble performance across threshold strategies with FAR-based classification on O3 GraceDB public events. The neural network consistently provides markedly improved identification of terrestrial (retracted) events regardless of the threshold used.}
\label{fig:gracedb_confusion_matrices}
\end{figure*}

\begin{figure*}[htbp]
\centering
\includegraphics[width=0.99\textwidth]{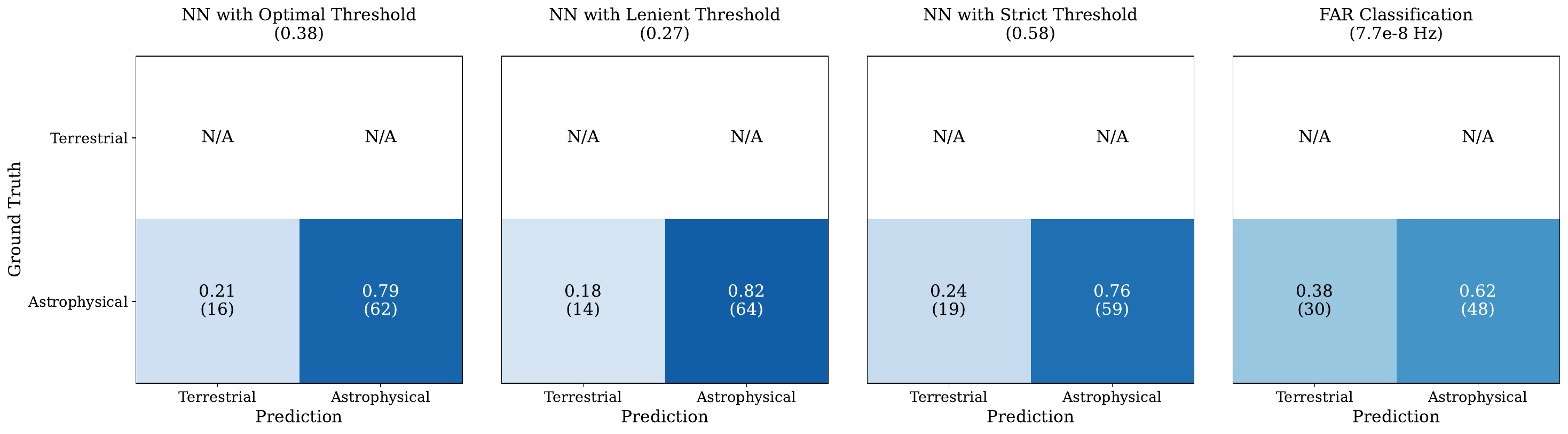}
\caption{Confusion matrices comparing neural-network (NN) ensemble performance under varying threshold strategies with FAR-based classification for the O3 GWTC-3 confident events catalog~\cite{PhysRevX.11.021053,PhysRevX.13.041039}. Terrestrial panels are NaN, as all GWTC-3 confident events are expected to be genuine astrophysical signals. The neural-network ensemble consistently outperforms FAR-based classification in the astrophysical category, independent of the threshold applied.}
\label{fig:gwtc_confusion_matrices}
\end{figure*}

In Fig.~\ref{fig:gracedb_confusion_matrices} and Fig. ~\ref{fig:gwtc_confusion_matrices} we highlight the performance of our neural network ensemble across different threshold strategies when applied to GraceDB public events and GWTC-3 confident events~\cite{PhysRevX.11.021053,PhysRevX.13.041039}.

For GraceDB public astronomical alerts~\footnote{https://gracedb.ligo.org/superevents/public/}, which include both high-significance signals and subsequently retracted terrestrial events, our optimal threshold (0.38) achieves 89\% sensitivity while correctly identifying 8 of 21 terrestrial retracted events as non-astrophysical. The lenient threshold (0.27) increases sensitivity to 91\% but incorrectly classifies 15 of 21 terrestrial retracted events as astrophysical, whereas the strict threshold (0.58) reduces this number to 11 of 21 but lowers sensitivity to 85\%.
The FAR-based classification reaches 95\% sensitivity but misclassifies 20 of 21 terrestrial retracted events, demonstrating poor separation between genuine signals and terrestrial contamination.

For GWTC-3 confident events, which represent the highest-confidence GW detections, as shown in Fig.~\ref{fig:gwtc_confusion_matrices}, our neural network shows consistent performance across all threshold strategies, maintaining 76-82\% sensitivity. We also took all the offline analysis $p_{\mathrm{astro}}$ values for the individual GWTC-3 confident events, 78 in total, and found that their mean $p_{\mathrm{astro}}$ is $0.9$, indicating some residual ambiguity even for the most extensive offline analyses. The FAR classification achieves 62\% sensitivity on GWTC-3 events, substantially lower than our neural network approaches, suggesting that our ensemble method can identify astrophysical signals that traditional statistical methods might miss or classify with lower confidence. This demonstrates the potential for machine learning approaches to enhance the detection efficiency of GW searches while maintaining reliability for real-time deployment.

A noteworthy result arises from the interaction between the ensemble model and the imperfect labeling inherent to real-time data. During training, the model had no a priori knowledge of which O3 GraceDB events would later be confirmed as astrophysical and included in the GWTC-3 catalogs~\cite{PhysRevX.11.021053,PhysRevX.13.041039}. Despite the structured filtering methodology described in Section ~\ref{sec:data}, 26 out of the 78 GWTC-3 confident events were included in the training noise class due to their high FAR value during the live run.
Crucially, when evaluated on the whole catalog of the 78 GWTC-3 confident events~\cite{PhysRevX.11.021053,PhysRevX.13.041039}, the ensemble correctly classified 13 of these 26 events as astrophysical, despite having been exposed to them as noise during training. This behavior demonstrates that the model does not simply memorize training labels, but instead learns a physically meaningful decision boundary that generalizes beyond imperfect or weak supervision. In effect, the ensemble is able to ``recover'' genuine signals even when their labels during training are contaminated by the limitations of real-time significance estimates.

This result has important implications for real-time deployment. In operating real-time and offline GW searches, definitive astrophysical labels are only assigned months later following extensive offline analyses. The demonstrated ability of the ensemble to identify true signals under label uncertainty indicates that it can act as a complementary real-time classifier, providing early, data-driven assessments of astrophysical origin that are robust to incomplete or evolving information. This robustness to label contamination, combined with its superior sensitivity relative to FAR-based methods, highlights the potential of out ML ensemble approach to enhance real-time detection efficiency and prioritization without compromising reliability.

\subsection{Outlier and Correlation Analysis}

\begin{figure*}[htbp]
\centering
\includegraphics[width=0.99\textwidth]{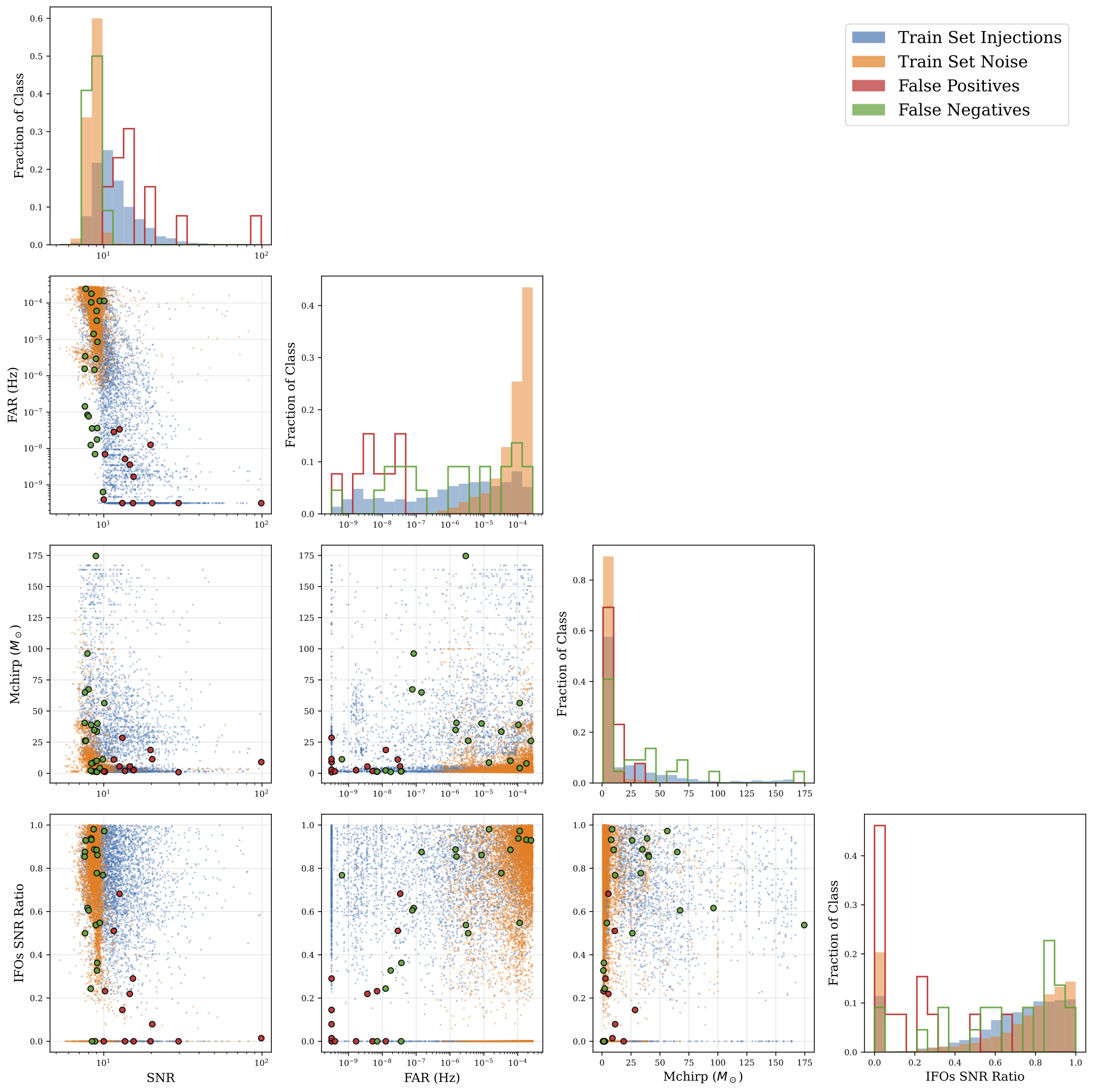}
\caption{Corner plot showing distributions and pairwise relationships of key features (SNR, FAR, Mchirp, IFOs SNR Ratio) for training set injections, noise, and misclassifications (false positives, false negatives).}
\label{fig:misclassification_analysis_corner_plot}
\end{figure*}

To better understand the origins of the misclassifications observed in the GraceDB high-significance public astronomical alerts and GWTC-3 confident event sets, we conducted a detailed outlier analysis aimed at characterizing the properties of false positives and false negatives relative to the ground truth training populations. This investigation focused on identifying systematic patterns in key features and locating these events within the broader feature space. In Figure ~\ref{fig:misclassification_analysis_corner_plot} misclassifications patterns suggest that false positives closely mimic the ground truth injection population, with distributions showing high SNR ($\gtrsim 10$), low FAR ($\lesssim 10^{-7}\,\mathrm{Hz}$), and low chirp mass ($\mathcal{M}_\mathrm{chirp} \lesssim 10\,M_\odot$). In nearly all feature-pair scatter plots, false positive points lie squarely within the injection-dominated regions (blue), explaining their promotion to the positive class. False negatives, in contrast, mimic the ground truth noise population, typically exhibiting low SNR ($\lesssim 10$), high FAR ($\gtrsim 10^{-6}\,\mathrm{Hz}$), and higher chirp mass ($\mathcal{M}_\mathrm{chirp} \gtrsim 10\,M_\odot$), clustering within the noise-dominated regions (orange) across the feature space. Correlation analysis between input features and ensemble outputs further highlights how chirp mass, SNR, and FAR shape both classification decisions and inter-model disagreement, providing guidance for future targeted retraining and feature refinement.

\begin{figure}[htbp]
\centering
\includegraphics[width=0.99\columnwidth]{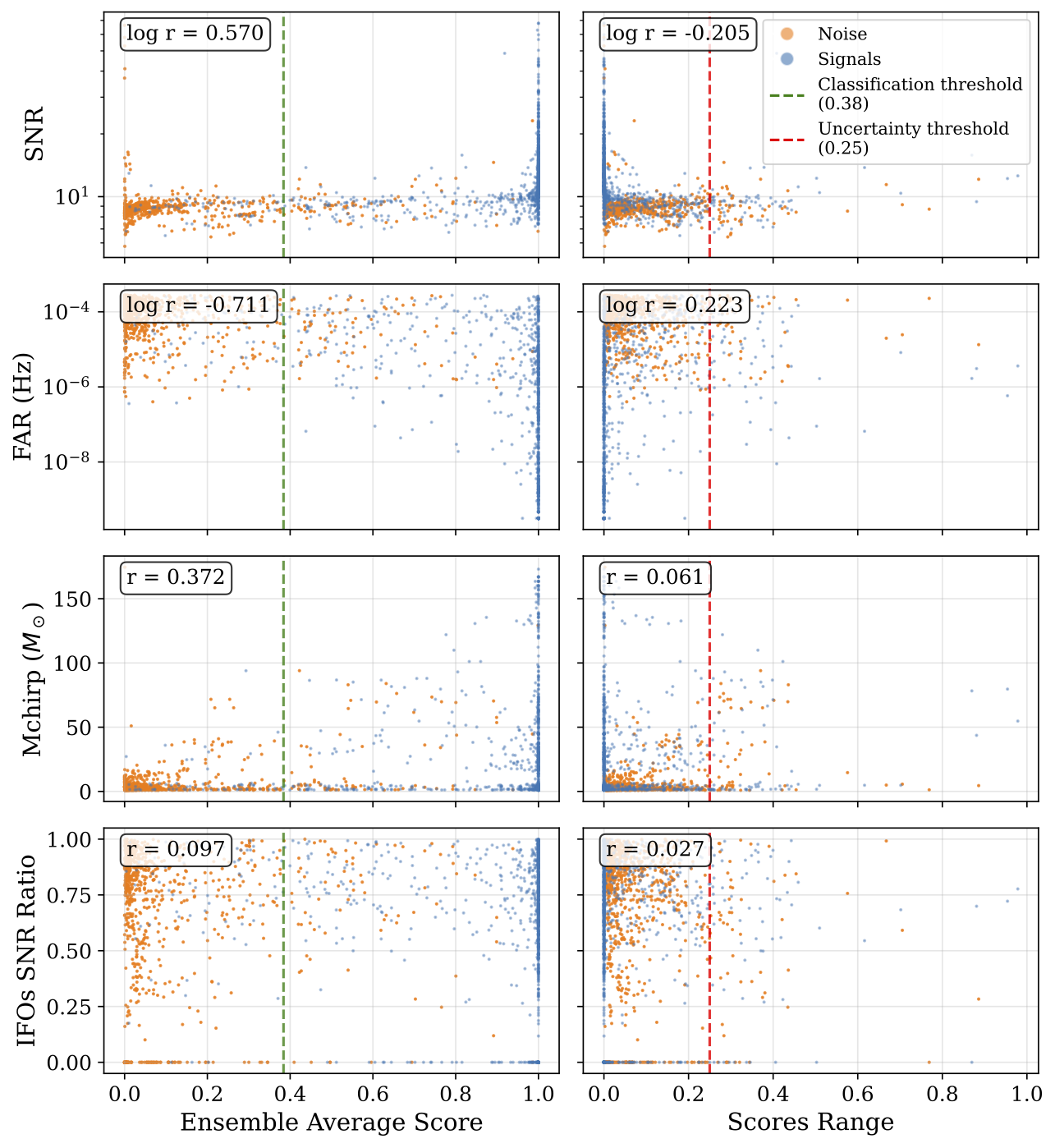}
\caption{Scatter plots showing correlations between input features and ensemble performance measures, with classification and uncertainty thresholds marked}
\label{fig:input_features_scatter_plots}
\end{figure}

\begin{figure}[htbp]
\centering
\includegraphics[width=0.99\columnwidth]{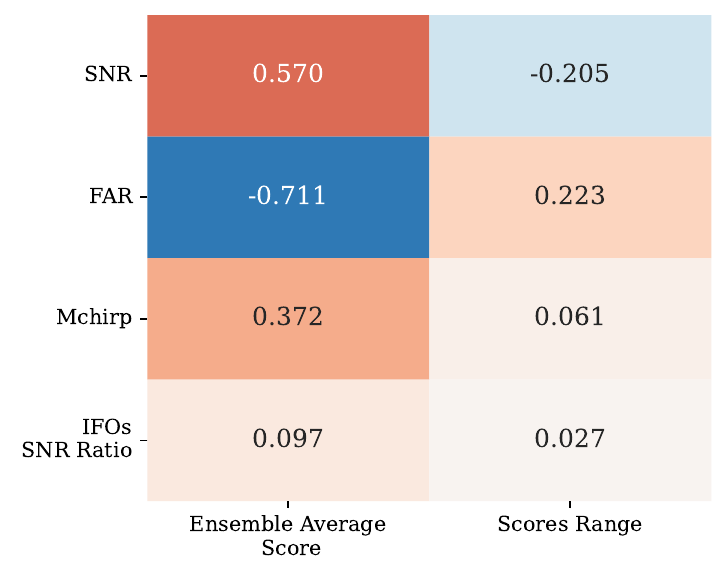}
\caption{Pearson correlation heatmap showing relationships between signal metrics and ensemble performance measures.}
\label{fig:pearson_correlation_heatmap}
\end{figure}

We conducted a correlation analysis to quantify the relationship between key input features and two ensemble-based classification metrics: the {average ensemble score} and the {score range} (model-to-model variation). This analysis was performed on the held-out test set, comprising 1000 MDC injections and 1000 O3 noise triggers. Figures ~\ref{fig:input_features_scatter_plots} and ~\ref{fig:pearson_correlation_heatmap} summarize the results. The scatter plots show how each feature, SNR, FAR, chirp mass ($\mathcal{M}_\mathrm{chirp}$), and interferometer SNR ratio, varies with both the average ensemble score (left panels) and the score range (right panels), with Pearson correlation coefficients annotated. The heatmap provides a compact view of these correlation values for all feature-metric pairs.

The results indicate that {SNR} is positively correlated with average score ($r=0.570$), meaning higher-SNR events tend to be assigned higher classification confidence, while its negative correlation with score range ($r=-0.205$) suggests reduced inter-model disagreement for strong signals. Conversely, {FAR} shows the strongest negative correlation with average score ($r=-0.711$), reflecting the model’s ability to output higher confidence scores for low-FAR candidates, and a moderate positive correlation with score range ($r=0.223$), indicating that events with higher FAR are more likely to yield uncertain or inconsistent predictions. Chirp mass exhibits a moderate positive correlation with an average score of ($r=0.372$) and negligible correlation with score range ($r=0.061$), suggesting that while higher chirp mass is mildly associated with higher confidence, it does not strongly influence inter-model disagreement. Finally, the {IFOs SNR ratio} shows weak correlations with both metrics ($r=0.097$ and $r=0.027$ respectively), implying limited influence on either classification confidence or uncertainty in the ensemble.  

Overall, these findings highlight that SNR and FAR are the dominant drivers of classification confidence, while FAR also contributes to model disagreement. These insights can guide targeted improvements in feature engineering, retraining, and ensemble calibration, particularly in regions of parameter space associated with low-confidence or inconsistent classifications.

\section{Conclusions} \label{sec:conclusion}

We have presented our Best-Of-All-Worlds (\texttt{BOAW}) pipeline, an ensemble-based neural network classifier designed to integrate outputs from multiple compact binary coalescence (CBC) search pipelines used in astrophysical searches by the LIGO-Virgo-KAGRA collaborations. By combining complementary information across search algorithms, the ensemble significantly improves signal vs. noise discrimination, achieving high recall with substantially reduced false positive rates compared to individual pipelines and their logical combinations.

Our results on the held-out test set comprising MDC injections~\cite{MDCPAPER} and O3 noise triggers demonstrate the effectiveness of this approach. At a 90\% true positive rate (TPR), the ensemble achieves an order-of-magnitude reduction in false positives relative to the union of pipelines, while maintaining broad coverage across all event types. The method also generalizes to GraceDB public alerts and GWTC-3 events, outperforming standard false alarm rate(FAR)-based thresholds in both sensitivity and specificity. Analyses of score distributions, confusion matrices, and correlation patterns confirm that the ensemble provides reliable classifications with principled uncertainty estimates. They also highlight the dominant role of SNR and FAR in shaping the algorithmic performance.

We note, however, that while results on the test set validate the model’s ability to separate injections from noise, performance on GraceDB significant and confident catalog events does not reach the same level. This reflects a few contributing factors, consistent with our misclassifications analysis (Fig.~\ref{fig:misclassification_analysis_corner_plot}) that includes the fact that (i) real astrophysical events differ in distribution from the MDC injections used in training, as shown by false positives clustering in injection-dominated regions and false negatives overlapping with noise regions; (ii) the current feature set, while informative, is limited, motivating exploration of additional search-pipeline data products that may encode richer astrophysical or instrumental information; and (iii) the MDC methodology, which injects simulated signals into archival O3 data at high rates, introduces an upward bias in FAR estimates when pipelines build their background models~\cite{MDCPAPER}, thereby reducing representativeness for true O3 GW events.

In this first analysis we relied on simulated and noise data from a 5-week period of LIGO-Virgo O3 run in order to train our algorithm which we then applied on the entire O3 dataset. 
This was mostly dictated by the availability of comprehensive MDC data that can be used for training.
Over extended observing runs of the GW instruments and with evolving detector sensitivities, changing noise environments, and updates to upstream search pipelines, monitoring of performance and algorithm retraining will be required.

The \texttt{BOAW} pipeline can be incorporated during real-time event analysis, complementing existing search pipelines in low-latency operations and addressing the trials factor issue that remains mostly unaddressed within the LIGO-Virgo-KAGRA public alert and catalog publications. Such integration could enable more reliable classification of candidate events, reduce the number of false alerts sent to the astronomical community, and improve prioritization for multi-messenger follow-up.
Taken together, these contributions show how machine learning ensembles can provide robust, flexible, and scalable enhancements to GW data analysis pipelines. By reducing false alerts, streamlining event classification, and adapting to evolving detector conditions, the \texttt{BOAW} framework represents a step toward more reliable real-time astrophysical alerts and more efficient use of multi-messenger follow-up resources.

\section{Acknowledgements} \label{ack}
NM, TM, DC and EK acknowledge support from the National Science Foundation (NSF) under award PHY-2309200 to the LIGO Laboratory and under PHY-2117997 to the NSF Institute on Accelerated AI Algorithms for Data Driven Discovery (A3D3, http://a3d3.ai/). 
This material is based upon work supported by NSF's LIGO Laboratory which is a major facility fully funded by the NSF.

\clearpage

\bibliography{references-final.bib} 
\bibliographystyle{apsrev}

\end{document}